\documentclass[11pt]{article}
\usepackage{graphicx}
\usepackage[T1]{fontenc}
\usepackage[utf8]{inputenc}
\usepackage{lmodern}
\usepackage{amsfonts}
\usepackage{amsmath}
\usepackage{amssymb}
\usepackage{amsthm}
\usepackage[numbers]{natbib}

\begin{document}

\title{A variationally consistent mesoscopic Cosserat theory with
distributed defects and configurational forces}
\author{Lev Steinberg}
\maketitle

\begin{abstract}
We develop a variationally consistent mesoscopic extension of Cosserat
elasticity motivated by the breakdown of compatibility in classical
formulations. By admitting compatibility-breaking perturbations, the
classical theory ceases to remain closed under admissible variations,
necessitating an enlargement of the constitutive framework. This leads
naturally to a formulation in which torsion and curvature are treated as
independent distributed measures of defects.

The theory is constructed using a Palatini-type variational approach, with
the coframe and connection as independent fields. The resulting
Euler--Lagrange equations yield both the standard balance laws and
defect-related excitation fields. Material invariance gives rise to
configurational forces and moments, which emerge as Noether currents and are
directly linked to defect transport governed by the Bianchi identities.

The framework provides a unified description of defect kinematics,
configurational mechanics, and microstructural evolution. Illustrative
examples and numerical evaluations demonstrate how defect transport
generates configurational forces and highlight the underlying Maxwell-type
structure of the theory. The proposed formulation offers a consistent
geometric foundation for the analysis of structured solids with evolving
internal geometry and provides a basis for future developments in defect
dynamics and dissipative processes.

\end{abstract}

\title{ \textbf{MSC}: 74A05, 74A15, 74E15, 74A30, 53C05}

\section{Introduction}

Cosserat (micropolar) elasticity extends classical continuum mechanics by
introducing independent microrotations and associated couple stresses. It
provides an effective framework for modeling size effects and materials with
internal structure  \cite{Nowacki1972}\cite{Eremeyev2012}\cite%
{Dyszlewicz2012}\cite{Eringen1999}In its classical form, however, the
kinematic fields are constrained by compatibility conditions corresponding
to vanishing torsion and curvature.

In many physically relevant situations, such as defect evolution,
localization phenomena, and microstructural rearrangements, these
compatibility conditions are violated. In particular, localization
processes, in which deformation concentrates in narrow regions, naturally
generate incompatibility. This viewpoint is consistent with configurational
mechanics, where incompatibility acts as a source of material forces \cite%
{Maugin1993}\cite{Gurtin2000}\cite{Eshelby1951}

Despite its generality, classical Cosserat elasticity \cite{Cosserat1909}%
\cite{Eringen1999}remains restricted to compatible configurations and
therefore ceases to be variationally closed once incompatibility is admitted
into the admissible class. In particular, the classical theory does not
penalize defect measures or their gradients, allowing localization to occur
with vanishing energetic cost in the limit. This leads to ill-posedness and
signals the breakdown of the classical description.

The present work addresses this deficiency by introducing a mesoscopic
extension of Cosserat elasticity, defined as the minimal constitutive
enlargement required to restore variational closure under loss of
compatibility. In this framework, torsion and curvature enter the
constitutive structure as distributed defect measures \cite{Kondo1952}\cite%
{Kroner1981}\cite{Yavari2012} This extension leads naturally to the
emergence of configurational forces \cite{Eshelby1951}\cite{Maugin1993} and
configurational couples as intrinsic features of the theory.

These configurational quantities admit a dual interpretation. On one hand,
they arise as Noether currents associated with material invariance of the
action. On the other hand, they appear as driving terms in the evolution of
defect measures governed by dynamic Bianchi identities \cite{Hehl1995}\cite%
{Roychowdhury2013}. This dual structure provides a unified variational and
geometric origin for defect-induced forces.

The geometric description of defects has a long history, originating in the
works of Kondo, Bilby, and Kr\"{o}ner, and further developed by Maugin,
Gurtin, Edelen, and others \cite{Kadic.Edelen1983},\cite{Kroner1981},\cite%
{deWit1973}\cite{Edelen 1985}. Earlier studies of strain singularities in
classical elasticity \cite{Steinberg} were based on a phenomenological
description of localized deformation. The present framework provides a
systematic variational and geometric formulation in which such behavior
emerges naturally from the governing equations.

Based on Cartan's theory of connections, torsion, and curvature \cite%
{Cartan1922}\cite{Cartan1967}\cite{Frankel}, these approaches represent
dislocations and disclinations as geometric incompatibilities. However, most
existing formulations do not employ a Palatini-type variational structure
with independent coframe and connection fields, nor do they fully exploit
the associated Noether framework.

The present mesoscopic Cosserat theory extends and builds upon a
Palatini-type variational formulation of classical Cosserat elasticity \cite%
{Cosserat1909},\cite{Marsden1983}developed in \cite{Steinberg2}, in which
the coframe and connection are treated as independent fields. The resulting
Euler--Lagrange equations yield both standard balance laws and
configurational balances \cite{Maugin1993}\cite{Gurtin2000} associated with
material invariance. A central result is the explicit decomposition of
configurational sources into stress--torsion, couple--curvature, and
higher-order incompatibility contributions.

The transition from classical to mesoscopic behavior is driven by increasing
external loading, which leads to loss of stability \cite{Timoshenko1985},%
\cite{Marsden1983} and the onset of distributed incompatibility.

\paragraph{Basic notation.}

We employ standard differential-geometric notation throughout. Latin indices 
$i,j,k$ refer to the internal (microstructural) frame, while capital indices 
$A,B,C$ denote material coordinates. The operator $D$ denotes the exterior
covariant derivative, and $\iota _{X}$ denotes contraction with a vector
field $X$. The coframe is denoted by $e^{i}$, the connection by $\omega
^{i}{}_{j}$, and the associated torsion and curvature by $T^{i}$ and $\Omega
^{i}{}_{j}$, respectively. The quantities $\Sigma _{i}$ and $M^{i}{}_{j}$
denote force and couple stresses, while $H_{i}$ and $G^{i}{}_{j}$ denote the
corresponding defect excitation fields.

\section{Geometric Setup}

Classical elasticity describes the deformation of a continuum in terms of
compatible strain fields, implicitly assuming the absence of internal
defects. However, at the mesoscopic level, and in particular near
instability or localization phenomena, this assumption breaks down. The
internal structure of the material may exhibit intrinsic incompatibility
that cannot be captured within a purely strain-based framework.

To account for this, we adopt a geometric formulation in which the
fundamental variables are a coframe field and an independent connection.
These fields describe, respectively, the translational and rotational
components of the internal material structure and allow for the presence of
distributed defects.

Let $M$ be a three-dimensional material body. We consider evolution over a
time interval $I \subset \mathbb{R}$, so that all fields depend on $(X,t)
\in M \times I$.

The internal structure of the material is described by a coframe field 
\begin{equation*}
e^{i}(X,t) \in \Omega^{1}(M),
\end{equation*}
which represents the translational part of the material microstructure.

The rotational part of the internal structure is described by an independent
connection 1-form 
\begin{equation*}
\omega^{i}{}_{j}(X,t) \in \Omega^{1}(M,so(3)).
\end{equation*}

The associated torsion and curvature are defined by 
\begin{equation*}
T^{i} = De^{i} = de^{i} + \omega^{i}{}_{k} \wedge e^{k} \in \Omega^{2}(M),
\end{equation*}
\begin{equation*}
\Omega^{i}{}_{j} = D\omega^{i}{}_{j} = d\omega^{i}{}_{j} + \omega^{i}{}_{k}
\wedge \omega^{k}{}_{j} \in \Omega^{2}(M,so(3)).
\end{equation*}

These quantities represent, respectively, the translational and rotational
incompatibility of the internal material structure. In the present
framework, torsion $T^{i}$ and curvature $\Omega ^{i}{}_{j}$ enter the
constitutive structure as distributed defect measures \cite{Kondo1952}\cite%
{Kroner1981}\cite{Yavari2012}.

This stands in contrast to classical Cosserat elasticity  \cite{Cosserat1909}%
\cite{Eringen1999}, where compatibility is enforced by 
\begin{equation*}
T^{i}=0,\qquad \Omega ^{i}{}_{j}=0,
\end{equation*}%
corresponding to the absence of defects.

If the connection is locally compatible, there exists a microrotation field $%
H(X) \in SO(3)$ such that 
\begin{equation*}
dH = H \omega,
\end{equation*}
and hence 
\begin{equation*}
\omega = H^{T} dH,
\end{equation*}
so that the connection is locally pure gauge \cite{Steinberg2}.

All fields are defined on the material manifold $M$ and are expressed in
material coordinates $X$.

\paragraph{Remark on the Microstructural (Fabric) Connection}

The geometric setting introduced above admits different interpretations
depending on the role assigned to the connection. In the present work, the
connection is not associated with the geometry of the ambient space, but is
instead interpreted as a microstructural (fabric) field describing the
internal rotational state of the material.

Accordingly, we distinguish between: (i) a geometric connection associated
with the ambient space, and (ii) a microstructural connection associated
with the internal rotational degrees of freedom of the material.

In classical differential geometry, a connection is typically induced by a
metric and reflects the intrinsic geometry of space. A standard example is
the Levi--Civita connection, uniquely determined by metric compatibility and
vanishing torsion, whose curvature measures the deviation of space from
Euclidean geometry.

In contrast, in the present theory the ambient space is assumed to remain
Euclidean. No non-Euclidean metric structure is introduced. Instead, the
connection is treated as an independent field describing the internal
rotational structure of the material.

Let 
\begin{equation*}
\omega \in \Omega^{1}(M,\mathfrak{so}(3))
\end{equation*}
be a connection 1-form. We interpret $\omega$ as a microstructural (fabric)
connection representing the local rotational state of material elements. It
is not derived from a metric and does not encode the geometry of space.

The associated curvature 
\begin{equation*}
\Omega = d\omega + \omega \wedge \omega \in \Omega^{2}(M,\mathfrak{so}(3))
\end{equation*}
measures the incompatibility of the microstructural rotational field and is
therefore interpreted as a density of rotational defects, that is, as a
disclination-type measure, rather than as curvature of the ambient space.

In particular, $\Omega = 0$ corresponds to rotational compatibility, in
which case the connection is locally integrable and admits a representation
in terms of a microrotation field $H(X)\in SO(3)$, 
\begin{equation*}
\omega = H^{T} dH.
\end{equation*}

The essential feature of the present formulation is thus the separation
between geometric and material structure: the ambient space remains
Euclidean, while nontrivial torsion and curvature arise from the internal
microstructure of the material. All fields are defined on the material
manifold $M$ and expressed in material coordinates $X$.

\section{Loss of Compatibility and Variational Closure}

\subsection{Classical Cosserat elasticity}

We consider a reduced classical Cosserat formulation in which the
independent fields are the coframe $e^{i}$ and the connection $%
\omega^{i}{}_{j}$, restricted to the compatible class. The stored energy
density is given by 
\begin{equation*}
W_{cl} = W_{cl}(e,\omega),
\end{equation*}
subject to the compatibility constraints 
\begin{equation*}
T^{i} = De^{i} = de^{i} + \omega^{i}{}_{k} \wedge e^{k} = 0,
\end{equation*}
\begin{equation*}
\Omega^{i}{}_{j} = D\omega^{i}{}_{j} = d\omega^{i}{}_{j} + \omega^{i}{}_{k}
\wedge \omega^{k}{}_{j} = 0.
\end{equation*}

The corresponding energy functional is 
\begin{equation*}
\Pi_{cl}[e,\omega;\Lambda] = \int_{\Omega_{0}} W_{cl}(e,\omega;\Lambda)\, dV,
\end{equation*}
where $\Lambda \in \mathbb{R}$ is a loading parameter.

\subsection{Stability and loss of coercivity}

Let $\eta = (\delta e, \delta \omega)$ denote an admissible perturbation
belonging to the compatible perturbation space $\mathcal{V}_{comp}$. The
second variation of the energy at equilibrium defines the quadratic form 
\begin{equation*}
Q_{\Lambda}(\eta) := \delta^{2}
\Pi_{cl}[e_{\Lambda},\omega_{\Lambda};\Lambda](\eta,\eta).
\end{equation*}

We define the lowest stability value 
\begin{equation*}
\mu_{1}(\Lambda) = \inf_{\eta \in \mathcal{V}_{comp} \setminus \{0\}} \frac{%
Q_{\Lambda}(\eta)}{\|\eta\|^{2}}.
\end{equation*}

Stability corresponds to $\mu_{1}(\Lambda) > 0$, while loss of stability
occurs at a critical load $\Lambda_{cr}$ such that 
\begin{equation*}
\mu_{1}(\Lambda_{cr}) = 0.
\end{equation*}

\subsection{Enlargement of admissible space}

To capture defect-like modes, one must enlarge the admissible class.
Consider a one-parameter family of perturbations 
\begin{equation*}
(e_{\tau},\omega_{\tau}) = (e,\omega) + \tau (\delta e, \delta \omega) +
o(\tau),
\end{equation*}
such that $(e,\omega)$ is compatible, while for $\tau \neq 0$ the perturbed
fields belong to an enlarged class $\mathcal{A}_{ext}$ satisfying 
\begin{equation*}
\delta T^{i} \neq 0 \quad \text{or} \quad \delta \Omega^{i}{}_{j} \neq 0.
\end{equation*}

\subsection{Loss of variational closure}

The classical energy density $W_{cl}(e,\omega)$ depends only on $(e,\omega)$
and does not include the incompatibility measures $(T,\Omega)$ in its
constitutive domain. Consequently, its first variation takes the form 
\begin{equation*}
\delta \Pi_{cl} = \int_{\Omega_{0}} \left( \frac{\partial W_{cl}}{\partial
e^{i}} \delta e^{i} + \frac{\partial W_{cl}}{\partial \omega^{i}{}_{j}}
\delta \omega^{i}{}_{j} \right) dV,
\end{equation*}
and cannot generate independent contributions associated with $\delta T^{i}$
or $\delta \Omega^{i}{}_{j}$.

Therefore, once compatibility-breaking perturbations are admitted, the
theory is no longer variationally closed.

\subsection{Main result}

\textbf{Theorem (Loss of variational closure).} Let the classical Cosserat
stored energy be given by 
\begin{equation*}
W_{cl} = W_{cl}(e,\omega),
\end{equation*}
with $(e,\omega)$ restricted to compatible configurations 
\begin{equation*}
T^{i} = 0, \qquad \Omega^{i}{}_{j} = 0.
\end{equation*}

Assume that at a critical load $\Lambda_{cr}$ there exist admissible
perturbations such that 
\begin{equation*}
\delta T^{i} \neq 0 \quad \text{or} \quad \delta \Omega^{i}{}_{j} \neq 0.
\end{equation*}

Then the classical Cosserat theory is not variationally closed with respect
to the enlarged admissible class.

Variational closure can be restored only by extending the constitutive
domain to include the incompatibility measures, 
\begin{equation*}
W_{mes}=W(\mathbf{e,\omega ,T,\Omega }),
\end{equation*}%
such that 
\begin{equation*}
W(\mathbf{e,\omega ,0,0})=W_{cl}(\mathbf{e,\omega }).
\end{equation*}

\textbf{Proof.} The classical energy depends only on $(e,\omega)$ and
therefore admits conjugate fields only for these variables. In the enlarged
admissible class, the perturbations generate independent variations $\delta
T^{i}$ and $\delta \Omega^{i}{}_{j}$, which are not associated with any
conjugate quantities in the classical theory.

Hence, the first variation cannot distinguish independent incompatibility
directions and is not sufficient to characterize stationarity with respect
to all admissible perturbations. This lack of conjugate fields implies loss
of variational closure.

\section{Mesoscopic Extension of Cosserat Elasticity}

\subsection{Constitutive enlargement}

To restore closure, the constitutive domain must be enlarged to include the
incompatibility measures. We introduce a mesoscopic stored energy of the
form 
\begin{equation*}
W_{mes} = W(e,\omega,T,\Omega),
\end{equation*}
where 
\begin{equation*}
T^{i} = De^{i}, \qquad \Omega^{i}{}_{j} = D\omega^{i}{}_{j}.
\end{equation*}

Consistency with the classical theory requires that 
\begin{equation*}
W(e,\omega,0,0) = W_{cl}(e,\omega).
\end{equation*}

\subsection{Minimal extension}

A minimal extension is given by 
\begin{equation*}
W_{mes}(e,\omega,T,\Omega) = W_{cl}(e,\omega) + \varepsilon
\Psi(e,\omega,T,\Omega),
\end{equation*}
with 
\begin{equation*}
\Psi(e,\omega,0,0) = 0.
\end{equation*}

\subsection{Conjugate fields}

The variational derivatives define 
\begin{equation*}
\Sigma_{i} = \frac{\partial W_{mes}}{\partial e^{i}}, \quad M^{i}{}_{j} = 
\frac{\partial W_{mes}}{\partial \omega^{j}{}_{i}},
\end{equation*}
\begin{equation*}
H_{i} = \frac{\partial W_{mes}}{\partial T^{i}}, \quad O^{i}{}_{j} = \frac{%
\partial W_{mes}}{\partial \Omega^{j}{}_{i}}.
\end{equation*}

\subsection{Variational completeness}

The first variation reads 
\begin{equation*}
\delta W_{mes} = \Sigma_{i} \, \delta e^{i} + M^{i}{}_{j} \, \delta
\omega^{j}{}_{i} + H_{i} \, \delta T^{i} + O^{i}{}_{j} \, \delta
\Omega^{j}{}_{i}.
\end{equation*}

This establishes variational completeness of the mesoscopic formulation.

\section{Lagrangian Formulation and Euler--Lagrange Equations}

\subsection{Variational setting}

We adopt a Palatini-type formulation \cite{Hehl1995} in which the coframe $%
e^{i}$ and the connection $\omega^{i}{}_{j}$ are treated as independent
fields. The action functional is defined by 
\begin{equation*}
\mathcal{A}[e,\omega] = \int_{I} \int_{B} L,
\end{equation*}
where $L$ is a Lagrangian density depending on the fields and their
kinematically induced quantities, 
\begin{equation*}
L = L(e^{i},\omega^{i}{}_{j},T^{i},\Omega^{i}{}_{j},J^{i},K^{i}{}_{j}).
\end{equation*}

The kinematic quantities are given by 
\begin{equation*}
T^{i} = De^{i}, \qquad \Omega^{i}{}_{j} = D\omega^{i}{}_{j},
\end{equation*}
\begin{equation*}
J^{i} = \partial_{t} e^{i}, \qquad K^{i}{}_{j} = \partial_{t}
\omega^{i}{}_{j}.
\end{equation*}

\subsection{Conjugate fields}

The variational derivatives of the Lagrangian define the conjugate
quantities 
\begin{equation*}
\Sigma_{i} = \frac{\partial L}{\partial e^{i}}, \qquad M^{i}{}_{j} = \frac{%
\partial L}{\partial \omega^{j}{}_{i}},
\end{equation*}
\begin{equation*}
H_{i} = \frac{\partial L}{\partial T^{i}}, \qquad O^{i}{}_{j} = \frac{%
\partial L}{\partial \Omega^{j}{}_{i}},
\end{equation*}
\begin{equation*}
P_{i} = \frac{\partial L}{\partial J^{i}}, \qquad Q^{i}{}_{j} = \frac{%
\partial L}{\partial K^{j}{}_{i}}.
\end{equation*}

\subsection{First variation}

The first variation of the Lagrangian takes the form 
\begin{equation*}
\delta L = \Sigma_{i} \wedge \delta e^{i} + M^{i}{}_{j} \wedge \delta
\omega^{j}{}_{i} + P_{i} \wedge \delta J^{i} + Q^{i}{}_{j} \wedge \delta
K^{j}{}_{i} + H_{i} \wedge \delta T^{i} + O^{i}{}_{j} \wedge \delta
\Omega^{j}{}_{i}.
\end{equation*}

The variations of the kinematic quantities are 
\begin{equation*}
\delta T^{i} = D(\delta e^{i}) + \delta \omega^{i}{}_{k} \wedge e^{k},
\end{equation*}
\begin{equation*}
\delta \Omega^{i}{}_{j} = D(\delta \omega^{i}{}_{j}),
\end{equation*}
\begin{equation*}
\delta J^{i} = \partial_{t} (\delta e^{i}), \qquad \delta K^{i}{}_{j} =
\partial_{t} (\delta \omega^{i}{}_{j}).
\end{equation*}

Substituting these relations into $\delta L$ and performing integration by
parts in time and covariant integration by parts in space, we obtain 
\begin{equation*}
\delta L = \delta e^{i} \wedge \left( \Sigma_{i} - \partial_{t} P_{i} - D
H_{i} \right) + \delta \omega^{j}{}_{i} \wedge \left( M^{i}{}_{j} -
\partial_{t} Q^{i}{}_{j} - D O^{i}{}_{j} - e^{i} \wedge H_{j} \right),
\end{equation*}
up to boundary terms.

\subsection{Euler--Lagrange equations}

Since the variations $\delta e^{i}$ and $\delta \omega ^{i}{}_{j}$ are
arbitrary, stationarity of the action yields the governing equations 
\begin{equation*}
DH_{i}+\partial _{t}P_{i}=\Sigma _{i},
\end{equation*}%
\begin{equation*}
DO^{i}{}_{j}+\partial _{t}Q^{i}{}_{j}+e^{i}\wedge H_{j}=M^{i}{}_{j}.
\end{equation*}

\subsection{Special Cases}

\paragraph{Stress-free case}

In the absence of stresses, 
\begin{equation*}
\Sigma _{i}=0,\qquad M^{i}{}_{j}=0,
\end{equation*}%
the system reduces to 
\begin{equation*}
DH_{i}+\partial _{t}P_{i}=0,
\end{equation*}%
\begin{equation*}
DO^{i}{}_{j}+\partial _{t}Q^{i}{}_{j}+e^{i}\wedge H_{j}=0.
\end{equation*}

This shows that defect excitations may persist even when observable stresses
vanish. In this regime, the excitation 1-forms $H_{i}$ and $O^{i}{}_{j}$ may
still be present, but they do not generate nonzero observable force-stress
or couple-stress fields. In this sense, the defective body is stress-free at
the level of the observable fields.

\paragraph{Static Case}

In the absence of inertia effects, the momenta vanish, 
\begin{equation*}
P_{i}=0,\qquad Q^{i}{}_{j}=0,
\end{equation*}%
and the equations reduce to 
\begin{equation*}
DH_{i}=\Sigma _{i},\qquad DO^{i}{}_{j}+e^{i}\wedge H_{j}=M^{i}{}_{j}.
\end{equation*}

\paragraph{Limiting Case}

In the compatible limit, 
\begin{equation*}
T^{i}=0,\qquad \Omega ^{i}{}_{j}=0,
\end{equation*}%
the defect excitations vanish, 
\begin{equation*}
H_{i}=0,\qquad O^{i}{}_{j}=0,
\end{equation*}%
and the classical Cosserat balance laws are recovered.

\subsection{Dissipative Extension}

\paragraph{Linear Extension}

We postulate the dissipative extension 
\begin{equation*}
D H_i+\partial_t P_i+\Sigma_i^{\mathrm{dis}}=\Sigma_i,
\end{equation*}
\begin{equation*}
D O^{i}{}_{j}+\partial_t Q^{i}{}_{j}+e^{i}\wedge H_j+M^{\mathrm{dis}%
\,i}{}_{j}=M^{i}{}_{j},
\end{equation*}
where 
\begin{equation*}
\Sigma_i^{\mathrm{dis}}=\gamma_T J_i, \qquad M^{\mathrm{dis}%
\,i}{}_{j}=\gamma_R K^{i}{}_{j},
\end{equation*}
with 
\begin{equation*}
\gamma_T>0, \qquad \gamma_R>0,
\end{equation*}
are translational and rotational defect-viscosity coefficients.

\paragraph{Energy balance}

Let $E$ denote the total energy density of the conservative sector,
including the kinetic contributions and the stored defect energy. Formally,
the dissipative equations lead to an energy balance of the form 
\begin{equation*}
\partial_t E + D = P_{\mathrm{ext}} - R,
\end{equation*}
where $P_{\mathrm{ext}}$ is the external power input, $D$ collects
flux-divergence terms, and the dissipation density is 
\begin{equation*}
R=\gamma_T \langle J,J\rangle+\gamma_R \langle K,K\rangle \ge 0.
\end{equation*}

Hence the added dissipative current terms produce genuine irreversible
decay. In particular, they are not equivalent to a mere modification of the
constitutive coefficients in $P_{i}$ and $Q^{i}{}_{j}$, since they generate
nonnegative dissipation proportional to $\Vert J\Vert ^{2}$ and $\Vert
K\Vert ^{2}$.

\section*{Maxwell-Type Structure of Mesoscopic Cosserat Elasticity}

The mesoscopic Cosserat theory admits a field-theoretic structure closely
analogous to electromagnetism when formulated in terms of differential
forms. This analogy is not merely suggestive, but structural, and appears at
the level of kinematics, variational principles, and balance laws.

\subsection{Homogeneous structure and Bianchi identities }

For clarity, we recall the kinematic relations, now interpreted as the
homogeneous sector of the field theory.

\begin{equation*}
T^{i}=De^{i}, \qquad \Omega^{i}{}_{j}=D\omega^{i}{}_{j}.
\end{equation*}

They satisfy the Bianchi identities\cite{Hehl1995}\cite{Roychowdhury2013}
\begin{equation*}
DT^{i}=\Omega ^{i}{}_{j}\wedge e^{j},\qquad D\Omega ^{i}{}_{j}=0.
\end{equation*}

These relations are geometric identities. They follow directly from the
definition of torsion and curvature and from the properties of the covariant
derivative. They do not depend on constitutive assumptions or on the choice
of Lagrangian. In this sense, they play the role of homogeneous field
equations.

Their dynamic counterparts 
\begin{equation*}
\partial_{t}T^{i}=D J^{i}+K^{i}{}_{j}\wedge e^{j}, \qquad
\partial_{t}\Omega^{i}{}_{j}=D K^{i}{}_{j},
\end{equation*}
describe the admissible transport of defect measures induced by the
evolution of the basic fields. These equations define admissible kinematic
evolution and do not, by themselves, determine the dynamics.

\subsection{Inhomogeneous structure and Euler--Lagrange equations}

The variational structure introduces the excitation fields 
\begin{equation*}
H_{i}=\frac{\partial L}{\partial T^{i}}, \qquad O^{i}{}_{j}=\frac{\partial L%
}{\partial \Omega^{j}{}_{i}},
\end{equation*}
and the momenta 
\begin{equation*}
P_{i}=\frac{\partial L}{\partial J^{i}}, \qquad Q^{i}{}_{j}=\frac{\partial L%
}{\partial K^{j}{}_{i}}.
\end{equation*}

The Euler--Lagrange equations take the form 
\begin{equation*}
D H_{i}+\partial_{t}P_{i}=\Sigma_{i},
\end{equation*}
\begin{equation*}
D O^{i}{}_{j}+\partial_{t}Q^{i}{}_{j}+e^{i}\wedge H_{j}=M^{i}{}_{j}.
\end{equation*}

These equations represent inhomogeneous balance laws in which the divergence
of the excitation fields is driven by the source terms $\Sigma_{i}$ and $%
M^{i}{}_{j}$.

\subsection{Structural correspondence with electromagnetism}

In classical electromagnetism, one introduces a potential 1-form $A$ and
defines the field strength 
\begin{equation*}
F = dA,
\end{equation*}
which satisfies the identity 
\begin{equation*}
dF = 0.
\end{equation*}

The excitation field is defined variationally as 
\begin{equation*}
H = \frac{\partial L}{\partial F},
\end{equation*}
and satisfies the inhomogeneous equation 
\begin{equation*}
dH = J.
\end{equation*}

In mesoscopic Cosserat theory, the fields $(e^{i},\omega^{i}{}_{j})$
generate the defect measures $(T^{i},\Omega^{i}{}_{j})$. The Bianchi
identities play the role of homogeneous equations, while the excitation
fields $(H_{i},O^{i}{}_{j})$ satisfy inhomogeneous Euler--Lagrange balances
driven by the stress sources $(\Sigma_{i},M^{i}{}_{j})$.

Thus, torsion and curvature act as field strengths, excitations arise
variationally, and stresses act as sources.

\subsection{Exact structural correspondence}

The analogy between electromagnetism and mesoscopic Cosserat theory may be
summarized as follows.

In electromagnetism, the fundamental variable is a potential $A$, the field
strength is $F=dA$, the homogeneous equation is $dF=0$, the excitation is $%
H=\partial L/\partial F$, and the inhomogeneous equation is $dH=J$.

In mesoscopic Cosserat theory, the basic fields $(e^{i},\omega ^{i}{}_{j})$
generate defect measures $(T^{i},\Omega ^{i}{}_{j})$, the Bianchi identities 
\cite{Hehl1995}\cite{Roychowdhury2013}  play the role of homogeneous
equations, the excitations are $(H_{i},O^{i}{}_{j})$, and the
Euler--Lagrange equations provide the inhomogeneous balances driven by $%
(\Sigma _{i},M^{i}{}_{j})$.

This correspondence shows that torsion and curvature play the role of field
strengths, while $(H_{i},O^{i}{}_{j})$ act as excitation variables, and $%
(\Sigma_{i},M^{i}{}_{j})$ act as sources.

\subsection{Excitations versus configurational forces }

The excitation fields $(H_{i},O^{i}{}_{j})$ are conjugate to the defect
measures and belong to the inhomogeneous field equations. However, they are
not configurational forces \cite{Eshelby1951}\cite{Maugin1993}
Configurational forces and couples arise at the Noether level, through
contraction of the balance laws with material generators and use of the
Bianchi identities \cite{Hehl1995}\cite{Roychowdhury2013}.

\section{Differences from classical Maxwell theory}

The mesoscopic Cosserat theory involves two field strengths, intrinsic
nonlinearity through the connection, and internal defect measures rather
than externally imposed fields.

\subsection{Maxwell-type interpretation of dissipation}

\begin{equation*}
\partial_{t}P_{i}+\gamma_{T}J_{i}, \qquad
\partial_{t}Q^{i}{}_{j}+\gamma_{R}K^{i}{}_{j}.
\end{equation*}

These terms are analogous to conductive currents in electromagnetism.

\subsection{Emergence of wave propagation}

Wave behavior emerges only after coupling the dynamic Bianchi identities 
\cite{Hehl1995}\cite{Roychowdhury2013} with the Euler--Lagrange equations
and constitutive relations.

In a linearized setting with constitutive laws 
\begin{equation*}
H_{i}=A_{ij}\ast T^{j},\qquad P_{i}=\rho _{ij}\ast J^{j},
\end{equation*}%
, one obtains schematically 
\begin{equation*}
\rho \,\partial _{t}^{2}J_{i}-a\,\Delta J_{i}=0,
\end{equation*}%
which represents wave propagation.

\subsection{Conclusion}

The mesoscopic Cosserat theory exhibits a Maxwell-type structure in which
Bianchi identities \cite{Hehl1995}\cite{Roychowdhury2013}define the
homogeneous kinematic sector, while Euler--Lagrange equations define the
inhomogeneous dynamic sector.

\section{Configurational Forces and Couples as Noether Currents}

\subsection{Variational setting and material invariance}

Let the action be 
\begin{equation*}
\mathcal{A}[e,\omega] = \int_{I}\int_{B} L,
\end{equation*}
with Lagrangian density 
\begin{equation*}
L = L(e^{i},\omega^{i}{}_{j},T^{i},\Omega^{i}{}_{j},J^{i},K^{i}{}_{j}).
\end{equation*}

Configurational balances arise from invariance of the action under localized
material transformations $X \mapsto X + \varphi \xi$, where $\varphi$ has
compact support. The Noether condition reads 
\begin{equation*}
\delta_{\varphi \xi} \mathcal{A} = 0.
\end{equation*}

\subsection{Localized material translations}

Let $E_{A}$ be a material basis vector. The induced variations are 
\begin{equation*}
\delta _{A}e^{i}=L_{\varphi E_{A}}^{D}e^{i},\qquad \delta _{A}\omega
^{i}{}_{j}=L_{\varphi E_{A}}^{D}\omega ^{i}{}_{j}.
\end{equation*}

Using the covariant Cartan formulas, 
\begin{equation*}
L^{D}_{\varphi E_{A}} e^{i} = \varphi \iota_{E_{A}} T^{i} + D(\varphi
\iota_{E_{A}} e^{i}),
\end{equation*}
\begin{equation*}
L^{D}_{\varphi E_{A}} \omega^{i}{}_{j} = \varphi \iota_{E_{A}}
\Omega^{i}{}_{j} + D(\varphi \iota_{E_{A}} \omega^{i}{}_{j}).
\end{equation*}

The time derivatives take the form 
\begin{equation*}
\partial_{t}(L^{D}_{\varphi E_{A}} e^{i}) =
(\partial_{t}\varphi)\,\iota_{E_{A}} e^{i} + \varphi \,\iota_{E_{A}} J^{i} +
\cdots,
\end{equation*}
\begin{equation*}
\partial_{t}(L^{D}_{\varphi E_{A}} \omega^{i}{}_{j}) =
(\partial_{t}\varphi)\,\iota_{E_{A}} \omega^{i}{}_{j} + \varphi
\,\iota_{E_{A}} K^{i}{}_{j} + \cdots,
\end{equation*}
where omitted terms are covariant transport contributions absorbed into the
current.

Substituting into the first variation and using the Euler--Lagrange
equations, we obtain 
\begin{equation*}
\delta _{A}L=D(\varphi S_{A}^{ext})+\partial _{t}(\varphi \Pi _{A}),
\end{equation*}

where the extended translational current is 
\begin{equation*}
S_{A}^{\mathrm{ext}}=\Sigma _{i}\,\iota _{E_{A}}e^{i}+M^{i}{}_{j}\,\iota
_{E_{A}}\omega ^{j}{}_{i}+H_{i}\wedge \iota _{E_{A}}T^{i}+O^{i}{}_{j}\wedge
\iota _{E_{A}}\Omega ^{j}{}_{i}-\iota _{E_{A}}L,
\end{equation*}%
and the configurational momentum is 
\begin{equation*}
\Pi _{A}=P_{i}\,\iota _{E_{A}}e^{i}+Q^{i}{}_{j}\,\iota _{E_{A}}\omega
^{j}{}_{i}.
\end{equation*}%
Since the action is invariant under localized material translations, 
\begin{equation*}
0=\delta _{A}\mathcal{A}=\int dt\int_{\mathcal{B}}\left( D(\varphi S_{A}^{%
\mathrm{ext}})+\partial _{t}(\varphi \Pi _{A})\right) .
\end{equation*}

Integrating by parts and using arbitrariness of $\varphi $ gives the local
Noether identity 
\begin{equation*}
DS_{A}^{ext}+\partial _{t}\Pi _{A}=0.
\end{equation*}

Splitting 
\begin{equation*}
S_{A}^{ext}=S_{A}+\Phi _{A},
\end{equation*}%
\begin{eqnarray*}
S_{A} &=&\Sigma _{i}\,\iota _{E_{A}}e^{i}+M^{i}{}_{j}\,\iota _{E_{A}}\omega
^{j}{}_{i}-\iota _{E_{A}}L, \\
\Phi _{A} &=&H_{i}\wedge \iota _{E_{A}}T^{i}+O^{i}{}_{j}\wedge \iota
_{E_{A}}\Omega ^{j}{}_{i},
\end{eqnarray*}%
Substituting this into the extended balance yields 
\begin{equation*}
DS_{A}+\partial _{t}\Pi _{A}=-D\Phi _{A}.
\end{equation*}

Now compute $D\Phi_{A}$ using the graded Leibniz rule: 
\begin{equation*}
D\Phi_{A} = DH_{i}\wedge \iota_{E_{A}}T^{i} - H_{i}\wedge
D(\iota_{E_{A}}T^{i}) + DG^{i}{}_{j}\wedge \iota_{E_{A}}\Omega^{j}{}_{i} -
G^{i}{}_{j}\wedge D(\iota_{E_{A}}\Omega^{j}{}_{i}).
\end{equation*}

Insert the Euler--Lagrange equations 
\begin{equation*}
DH_{i}=\Sigma_{i}-\partial_{t}P_{i}, \qquad
DG^{i}{}_{j}=M^{i}{}_{j}-\partial_{t}Q^{i}{}_{j}-e^{i}\wedge H_{j},
\end{equation*}
and then use the covariant Cartan identities together with the Bianchi
identities 
\begin{equation*}
D(\iota_{E_{A}}T^{i}) = \mathcal{L}^{D}_{E_{A}}T^{i} -
\iota_{E_{A}}(DT^{i}), \qquad D(\iota_{E_{A}}\Omega^{i}{}_{j}) = \mathcal{L}%
^{D}_{E_{A}}\Omega^{i}{}_{j} - \iota_{E_{A}}(D\Omega^{i}{}_{j}),
\end{equation*}
\begin{equation*}
DT^{i}=\Omega^{i}{}_{j}\wedge e^{j}, \qquad D\Omega^{i}{}_{j}=0.
\end{equation*}

After expansion, the mixed terms cancel exactly and one obtains 
\begin{equation*}
D\Phi _{A}=\iota _{E_{A}}T^{i}\wedge \Sigma _{i}+\iota _{E_{A}}\Omega
^{i}{}_{j}\wedge M^{j}{}_{i},
\end{equation*}%
The equations take place%
\begin{equation*}
DS_{A}+\partial _{t}\Pi _{A}=R_{A},
\end{equation*}%
with 
\begin{equation*}
R_{A}=\iota _{E_{A}}T^{i}\wedge \Sigma _{i}+\iota _{E_{A}}\Omega
^{i}{}_{j}\wedge M^{j}{}_{i}.
\end{equation*}

\subsection{Localized material rotations}

Now consider localized infinitesimal material rotations generated by 
\begin{equation*}
\varphi J_{AB}, \qquad J_{AB}=-J_{BA}.
\end{equation*}

The induced variations are 
\begin{equation*}
\delta_{AB} e^{i}=\mathcal{L}^{D}_{\varphi J_{AB}}e^{i}, \qquad
\delta_{AB}\omega^{i}{}_{j}=\mathcal{L}^{D}_{\varphi J_{AB}}\omega^{i}{}_{j}.
\end{equation*}

Proceeding exactly as in the translational case, but now with the rotational
generator, one inserts these variations into the first variation formula and
collects the coefficients of $D\varphi $ and $\partial _{t}\varphi $. On
shell this yields 
\begin{equation*}
\delta _{AB}L=D\!\left( \varphi M_{AB}^{\mathrm{ext}}\right) +\varphi \left(
E_{A}\wedge S_{B}^{\mathrm{ext}}-E_{B}\wedge S_{A}^{\mathrm{ext}}\right)
+\partial _{t}\!\left( \varphi \Pi _{AB}\right) ,
\end{equation*}%
where $M_{AB}^{\mathrm{ext}}$ is the extended rotational Noether current and 
$\Theta _{AB}$ is the corresponding temporal current.

Invariance of the action under localized material rotations therefore gives 
\begin{equation*}
0=\delta _{AB}\mathcal{A}=\int dt\int_{\mathcal{B}}\left[ D\!\left( \varphi
M_{AB}^{\mathrm{ext}}\right) +\varphi \left( E_{A}\wedge S_{B}^{\mathrm{ext}%
}-E_{B}\wedge S_{A}^{\mathrm{ext}}\right) +\partial _{t}\!\left( \varphi \Pi
_{AB}\right) \right] .
\end{equation*}

Integrating by parts and localizing we obtain the extended homogeneous
configurational moment balance.%
\begin{equation*}
DM_{AB}^{\mathrm{ext}}+E_{A}\wedge S_{B}^{\mathrm{ext}}-E_{B}\wedge S_{A}^{%
\mathrm{ext}}+\partial _{t}\Pi _{AB}=0.
\end{equation*}

Now we also write 
\begin{equation*}
M_{AB}^{\mathrm{ext}}=M_{AB}+\Psi _{AB},
\end{equation*}%
where $\Psi _{AB}$ is the defect-conjugate rotational contribution. Then the
extended balance becomes 
\begin{equation*}
DM_{AB}+E_{A}\wedge S_{B}-E_{B}\wedge S_{A}+\partial _{t}\Pi _{AB}=R_{AB},
\end{equation*}%
where%
\begin{equation*}
M_{AB}=\Sigma _{i}\,\iota _{J_{AB}}e^{i}+M^{i}{}_{j}\,\iota _{J_{AB}}\omega
^{j}{}_{i}-\iota _{J_{AB}}L,
\end{equation*}
\begin{equation*}
R_{AB}=-R_{BA}=\iota _{J_{AB}}T^{i}\wedge \Sigma _{i}+\iota _{J_{AB}}\Omega
^{i}{}_{j}\wedge M^{j}{}_{i},
\end{equation*}%
Thus, \ $J_{AB}$ be the generator of material rotations. with 
\begin{equation*}
\Pi _{AB}=P_{i}\,\iota _{J_{AB}}e^{i}+Q^{i}{}_{j}\,\iota _{J_{AB}}\omega
^{j}{}_{i}.
\end{equation*}

\subsection{Line defects and resultant configurational forces \protect\cite%
{Eshelby1951},\protect\cite{Maugin1993}}

Let $L\subset M$ be a defect line and let $\delta _{L}$ denote the Dirac
line measure supported on $L$. We assume that torsion and curvature
concentrate along $L$ according to 
\begin{equation*}
T^{A}=b^{A}\,\delta _{L},\qquad \Omega ^{AB}=\kappa ^{AB}\,\delta _{L},
\end{equation*}%
where $b^{A}$ is the Burgers vector and $\kappa ^{AB}=-\kappa ^{BA}$ is the
Frank-type disclination tensor.

\paragraph{Configurational force}

Let $\theta^{A}$ be the material coframe dual to $E_{A}$, 
\begin{equation*}
\iota_{E_{A}}\theta^{B} = \delta_{A}^{B}.
\end{equation*}

Expand 
\begin{equation*}
\Sigma_{B} = \Sigma_{BM}\theta^{M}, \qquad M^{CB} = M^{CB}_{M}\theta^{M},
\end{equation*}
so that 
\begin{equation*}
\iota_{E_{A}}\Sigma_{B} = \Sigma_{BA}, \qquad \iota_{E_{A}}M^{CB} =
M^{CB}_{A}.
\end{equation*}

Assume a line defect $L$ with concentrated measures 
\begin{equation*}
T^{A} = b^{A}\delta_{L}, \qquad \Omega^{AB} = \kappa^{AB}\delta_{L}.
\end{equation*}

The configurational force per unit length is 
\begin{equation*}
f_{A} = \int_{A_{\varepsilon}} R_{A},
\end{equation*}
which gives 
\begin{equation*}
f_{A} = b^{B}\Sigma_{BA} + \kappa^{BC} M^{CB}_{A}.
\end{equation*}

If $\kappa ^{AB}=0$, then 
\begin{equation*}
f_{A}=b^{B}\Sigma _{BA},
\end{equation*}%
which corresponds to the classical Peach--Koehler force.

When curvature is present, the additional term 
\begin{equation*}
\kappa ^{BC}M^{CB}{}_{A}
\end{equation*}%
provides a force contribution associated with the couple-stress field. This
term has no counterpart in classical elasticity and represents a genuinely
mesoscopic correction induced by disclination-type defects.

\paragraph{Configurational moment}

Let 
\begin{equation*}
J_{CD} = X_{C}E_{D} - X_{D}E_{C}.
\end{equation*}

Then for a 1-form $\alpha =\alpha _{M}\theta ^{M}$, 
\begin{equation*}
\iota _{J_{CD}}\alpha =X_{C}\alpha _{D}-X_{D}\alpha _{C}.
\end{equation*}%
Applying this to the force-stress and couple-stress fields gives 
\begin{equation*}
\iota _{J_{CD}}\Sigma _{A}=X_{C}\Sigma _{AD}-X_{D}\Sigma _{AC},
\end{equation*}%
\begin{equation*}
\iota _{J_{CD}}M^{BA}=X_{C}M^{BA}{}_{D}-X_{D}M^{BA}{}_{C}.
\end{equation*}%
The rotational configurational source is 
\begin{equation*}
R_{CD}=\iota _{J_{CD}}T^{A}\wedge \Sigma _{A}+\iota _{J_{CD}}\Omega
^{AB}\wedge M^{BA}.
\end{equation*}

The configurational moment per unit length is 
\begin{equation*}
m_{CD} = \int_{A_{\varepsilon}} R_{CD},
\end{equation*}
which yields 
\begin{equation*}
m_{CD} = b^{A}(X_{C}\Sigma_{AD} - X_{D}\Sigma_{AC}) +
\kappa^{AB}(X_{C}M^{BA}_{D} - X_{D}M^{BA}_{C}).
\end{equation*}

In three dimensions, define the axial vector 
\begin{equation*}
m_{E}=\frac{1}{2}\varepsilon _{E}^{\;CD}m_{CD}.
\end{equation*}

\subsection{Noether interpretation and relation to classical configurational
mechanics}

The preceding derivation can be summarized in a unified manner. \ The
configurational force and configurational couple balances arise as Noether
identities associated with localized material translations and rotations,
respectively. These identities hold provided that the Euler--Lagrange
equations are satisfied and the action is invariant under localized material
transformations. In this sense, configurational forces \cite{Eshelby1951}%
\cite{Maugin1993} and configurational couples are intrinsic consequences of
the variational structure and do not require separate postulates.

The present formulation extends the classical configurational mechanics of
Eshelby and Maugin in several essential directions. First, it provides a
fully covariant formulation in terms of differential forms. Second, it
incorporates torsion and curvature as distributed defect measures \cite%
{Kondo1952}\cite{Kroner1981}\cite{Yavari2012}Third, it yields both
configurational forces \cite{Eshelby1951}\cite{Maugin1993} and
configurational moments within a single Noether framework.

In this setting, the classical theory of Eshelby \cite{Eshelby1951} and its
extension by Maugin \cite{Maugin1993} appear as reduced limits. The tensor $%
S_{A}$ plays the role of an Eshelby-type configurational stress, while the
source term $R_{A}$ represents a distributed configurational force density
in the sense of Maugin.

The formulation is also consistent with the geometric theory of defects
developed by Kroner and Kondo, in which torsion and curvature represent
dislocations and disclinations. In contrast to classical singular
descriptions, the present theory treats these quantities as distributed
fields and embeds them into a unified variational and Noether framework.

This unified perspective shows that configurational forces \cite{Eshelby1951}%
\cite{Maugin1993}, defect transport, and material structure arise as
different manifestations of a single underlying geometric and variational
principle.

Thus, the configurational forces \cite{Eshelby1951}\cite{Maugin1993}arise
intrinsically from the variational structure of the mesoscopic theory. They
do not need to be postulated separately but emerge as a consequence of the
extended constitutive framework.This provides a unified origin for
Peach--Koehler-type forces and their curvature-driven generalizations.

\section{Dynamic Bianchi Identities}

The evolution equations follow from the time differentiation of the Bianchi
identities

\begin{equation*}
DT^{i}=\Omega ^{i}{}_{j}\wedge e^{j},\qquad D\Omega ^{i}{}_{j}=0,
\end{equation*}

\cite{Hehl1995}, together with the kinematic definitions of the rates. This
yields transport equations for torsion and curvature consistent with
geometric defect theory \cite{Yavari2012} \cite{Roychowdhury2013}

The geometric structure introduced above satisfies the (static) Bianchi
identities 
\begin{equation*}
DT^{i} = \Omega^{i}{}_{j} \wedge e^{j}, \qquad D\Omega^{i}{}_{j} = 0,
\end{equation*}
which express intrinsic geometric relations between torsion and curvature.

To describe the evolution of defect fields, we allow the coframe and
connection to depend on time. Introducing the kinematic rates 
\begin{equation*}
J^{i} := \partial_{t} e^{i}, \qquad K^{i}{}_{j} := \partial_{t}
\omega^{i}{}_{j},
\end{equation*}
we obtain the dynamic Bianchi identities by differentiating the defining
relations of torsion and curvature with respect to time.

For torsion, 
\begin{equation*}
\partial_{t} T^{i} = \partial_{t}(D e^{i}) = D(\partial_{t} e^{i}) +
(\partial_{t} \omega^{i}{}_{k}) \wedge e^{k},
\end{equation*}
which yields 
\begin{equation*}
\partial_{t} T^{i} = D J^{i} + K^{i}{}_{k} \wedge e^{k}.
\end{equation*}

Similarly, for curvature, 
\begin{equation*}
\partial_{t} \Omega^{i}{}_{j} = \partial_{t}(D \omega^{i}{}_{j}) =
D(\partial_{t} \omega^{i}{}_{j}),
\end{equation*}
so that 
\begin{equation*}
\partial_{t} \Omega^{i}{}_{j} = D K^{i}{}_{j}.
\end{equation*}

These relations govern the admissible evolution of torsion and curvature in
terms of the rates of the primary fields.

\textbf{Interpretation.} The dynamic Bianchi identities are purely
kinematic: they describe the transport of defect measures induced by the
evolution of the coframe and connection. They do not, by themselves,
determine the dynamical character of the evolution.

In this sense, they play a role analogous to homogeneous field equations in
gauge theories.

\textbf{Compatible (classical) limit.} In the compatible limit, 
\begin{equation*}
T^{i}=0,\qquad \Omega ^{i}{}_{j}=0,
\end{equation*}%
the defect measures vanish and the microstructure becomes locally
integrable. In this case, the mesoscopic contributions disappear and the
theory reduces to the classical defect-free Cosserat elasticity.

In particular, the defect conjugate quantities satisfy 
\begin{equation*}
H_{i} = 0, \qquad G^{i}{}_{j} = 0,
\end{equation*}
consistently with the absence of torsion and curvature in the constitutive
response.

Thus, the mesoscopic formulation recovers the classical Cosserat theory as a
special compatible case.

\section*{Complete System of Mesoscopic Elasticity Equations}

For clarity, we collect here the complete system of governing equations of
the mesoscopic Cosserat theory.

\subsection*{Field equations}

The Euler--Lagrange equations are 
\begin{equation*}
DH_{i}+\partial_{t}P_{i}=\Sigma_{i},
\end{equation*}
\begin{equation*}
DO^{i}{}_{j}+\partial_{t}Q^{i}{}_{j}+e^{i}\wedge H_{j}=M^{i}{}_{j}.
\end{equation*}

Here, $\Sigma_{i}$ and $M^{i}{}_{j}$ denote the generalized force stress and
couple stress conjugate to the fields $e^{i}$ and $\omega^{i}{}_{j}$,
respectively. The quantities $H_{i}$ and $O^{i}{}_{j}$ are defect stresses
or defect excitations conjugate to torsion and curvature. The quantities $%
P_{i}$ and $Q^{i}{}_{j}$ are the linear and angular momenta conjugate to the
rates $J^{i}$ and $K^{i}{}_{j}$.

\subsection*{Constitutive relations}

The constitutive quantities are determined from the Lagrangian $L$ by 
\begin{equation*}
\Sigma_{i}=\frac{\partial L}{\partial e^{i}}, \qquad M^{i}{}_{j}=\frac{%
\partial L}{\partial \omega^{j}{}_{i}},
\end{equation*}
\begin{equation*}
H_{i}=\frac{\partial L}{\partial T^{i}}, \qquad O^{i}{}_{j}=\frac{\partial L%
}{\partial \Omega^{j}{}_{i}},
\end{equation*}
\begin{equation*}
P_{i}=\frac{\partial L}{\partial J^{i}}, \qquad Q^{i}{}_{j}=\frac{\partial L%
}{\partial K^{j}{}_{i}}.
\end{equation*}

\subsection*{Geometrical setting}

The torsion and curvature are defined by 
\begin{equation*}
T^{i}=De^{i}=de^{i}+\omega^{i}{}_{k}\wedge e^{k},
\end{equation*}
\begin{equation*}
\Omega^{i}{}_{j}=D\omega^{i}{}_{j}=d\omega^{i}{}_{j}+\omega^{i}{}_{k}\wedge%
\omega^{k}{}_{j},
\end{equation*}
and the rates are 
\begin{equation*}
J^{i}=\partial_{t}e^{i}, \qquad K^{i}{}_{j}=\partial_{t}\omega^{i}{}_{j}.
\end{equation*}

\subsection*{Material configurational balances}

The configurational force and moment balances are 
\begin{equation*}
DS_{A}+\partial_{t}\Pi_{A}=R_{A},
\end{equation*}
\begin{equation*}
DM_{AB}+E_{A}\wedge S_{B}-E_{B}\wedge S_{A}+\partial_{t}\Pi_{AB}=R_{AB}.
\end{equation*}

The configurational stresses and configurational couple stresses are 
\begin{equation*}
S_{A}=\Sigma_{i}\,\iota_{E_{A}}e^{i}
+M^{i}{}_{j}\,\iota_{E_{A}}\omega^{j}{}_{i} -\iota_{E_{A}}L,
\end{equation*}
\begin{equation*}
R_{A}=\iota_{E_{A}}T^{i}\wedge\Sigma_{i}
+\iota_{E_{A}}\Omega^{i}{}_{j}\wedge M^{j}{}_{i},
\end{equation*}
\begin{equation*}
M_{AB}=\Sigma_{i}\,\iota_{J_{AB}}e^{i}
+M^{i}{}_{j}\,\iota_{J_{AB}}\omega^{j}{}_{i} -\iota_{J_{AB}}L,
\end{equation*}
\begin{equation*}
R_{AB}=-R_{BA} =\iota_{J_{AB}}T^{i}\wedge\Sigma_{i}
+\iota_{J_{AB}}\Omega^{i}{}_{j}\wedge M^{j}{}_{i}.
\end{equation*}

The configurational linear and angular momenta are 
\begin{equation*}
\Pi _{A}=P_{i}\,\iota _{E_{A}}e^{i}+Q^{i}{}_{j}\,\iota _{E_{A}}\omega
^{j}{}_{i},
\end{equation*}%
\begin{equation*}
\Pi _{AB}=P_{i}\,\iota _{J_{AB}}e^{i}+Q^{i}{}_{j}\,\iota _{J_{AB}}\omega
^{j}{}_{i}.
\end{equation*}

\bigskip

\subsection{Dynamic Bianchi transport}

The transport equations are 
\begin{equation*}
\partial_{t}T^{i}=DJ^{i}+K^{i}{}_{j}\wedge e^{j},
\end{equation*}
\begin{equation*}
\partial_{t}\Omega^{i}{}_{j}=DK^{i}{}_{j}.
\end{equation*}

\subsection*{Bianchi identities}

The kinematic identities are 
\begin{equation*}
DT^{i}=\Omega ^{i}{}_{k}\wedge e^{k},
\end{equation*}%
\begin{equation*}
D\Omega ^{i}{}_{j}=0.
\end{equation*}

\section{Illustrative Examples}

\subsection{2D Dynamic Bianchi Transport}

We consider a two-dimensional Cosserat-type setting with scalar connection $%
\omega $ and torsion and curvature represented as differential 2-forms. This
formulation is consistent with geometric and differential-form approaches to
defects in continua and Cosserat media  \  \cite{Kondo1952}\cite{Kroner1981}
\cite{Roychowdhury2013}\cite{Roychowdhury2017}\cite{Yavari2012} \cite%
{Yavari2012}\cite{NemethAdhikari2024}

The Hodge dual is taken with respect to the Euclidean area form, so that 
\begin{equation*}
\ast (dx\wedge dy)=1.
\end{equation*}%
The covariant exterior derivative is denoted by $D$.

\paragraph{Kinematic setting.}

Let the planar material strip be parameterized by $(x,y)$ with coframe 
\begin{equation*}
e^1 = dx, \qquad e^2 = dy,
\end{equation*}
and scalar micro-rotation connection 
\begin{equation*}
\omega = a(y,t)\,dx, \qquad a(y,t)=e^{-t}\sin(\pi y).
\end{equation*}
The rate of the connection is 
\begin{equation*}
K=\partial_t \omega = -e^{-t}\sin(\pi y)\,dx.
\end{equation*}

\paragraph{Defect fields.}

Since $de^{1}=0$ and $\omega \wedge e^{2}=a\,dx\wedge dy$, the torsion is 
\begin{equation*}
T^{1}=a\,dx\wedge dy=e^{-t}\sin (\pi y)\,dx\wedge dy,\qquad T^{2}=0.
\end{equation*}%
The curvature is 
\begin{equation*}
\Omega =d\omega =-\pi e^{-t}\cos (\pi y)\,dx\wedge dy.
\end{equation*}%
These definitions are standard in geometric defect theory, where torsion and
curvature represent distributed measures of dislocations and disclinations 
\cite{Kondo1952}\cite{Kroner1981}\cite{HehlObukhov2007}\cite{Yavari2012}

\paragraph{Defect excitations.}

With constitutive laws 
\begin{equation*}
H_1 = \mu_T *T^1, \qquad O = \mu_R *\Omega,
\end{equation*}
we obtain 
\begin{equation*}
H_1 = \mu_T e^{-t}\sin(\pi y), \qquad O = -\mu_R \pi e^{-t}\cos(\pi y).
\end{equation*}

\paragraph{Force stress.}

With $J^i=0$ and $P_i=0$, the balance reduces to 
\begin{equation*}
dH_1 = \Sigma_1,
\end{equation*}
hence 
\begin{equation*}
\Sigma_1 = \mu_T \pi e^{-t}\cos(\pi y)\,dy.
\end{equation*}

\paragraph{Couple stress.}

If 
\begin{equation*}
dO + \gamma_R K = M,
\end{equation*}
then 
\begin{equation*}
M = \mu_R \pi^2 e^{-t}\sin(\pi y)\,dy - \gamma_R e^{-t}\sin(\pi y)\,dx.
\end{equation*}

\paragraph{Dynamic Bianchi transport.}

The transport relations read 
\begin{equation*}
\partial _{t}T^{1}=K\wedge e^{2},\qquad \partial _{t}\Omega =dK.
\end{equation*}%
These are reduced forms of the dynamic Bianchi identities \cite{Hehl1995} 
\cite{Roychowdhury2013}governing defect transport \cite{Hehl1995} \cite%
{Roychowdhury2013} \cite{Fumeron2022}.

\paragraph{Interpretation.}

This example exhibits explicitly the chain 
\begin{equation*}
\omega \rightarrow (T,\Omega) \rightarrow (H,O) \rightarrow (\Sigma,M),
\end{equation*}
and shows that defect evolution is governed by the dynamic Bianchi
identities.

\subsection*{Time-Dependent Coframe and Connection}

\paragraph{Kinematic setting.}

Let 
\begin{equation*}
e^1 = b(y,t)\,dx, \qquad e^2=dy, \qquad b(y,t)=1+\varepsilon e^{-t}\sin(\pi
y),
\end{equation*}
and 
\begin{equation*}
\omega = a(y,t)\,dx, \qquad a(y,t)=a_0 e^{-t}\cos(\pi y),
\end{equation*}
where $a_0$ and $\varepsilon$ are constants.

\paragraph{Rates.}

The coframe and connection rates are 
\begin{equation*}
J^1 = \partial_t e^1 = -\varepsilon e^{-t}\sin(\pi y)\,dx, \qquad J^2=0,
\end{equation*}
and 
\begin{equation*}
K = \partial_t \omega = -a_0 e^{-t}\cos(\pi y)\,dx.
\end{equation*}

\paragraph{Defect fields.}

The torsion is 
\begin{equation*}
T^1 = de^1 + \omega \wedge e^2 = (a-b_y)\,dx \wedge dy,
\end{equation*}
with 
\begin{equation*}
b_y = \varepsilon \pi e^{-t}\cos(\pi y),
\end{equation*}
so that 
\begin{equation*}
T^1 = e^{-t}(a_0-\varepsilon \pi)\cos(\pi y)\,dx \wedge dy.
\end{equation*}
Also, 
\begin{equation*}
T^2=0.
\end{equation*}
The curvature is 
\begin{equation*}
\Omega = d\omega = a_0 \pi e^{-t}\sin(\pi y)\,dx \wedge dy.
\end{equation*}

\paragraph{Dynamic Bianchi identity.}

The torsion transport law is 
\begin{equation*}
\partial _{t}T^{1}=DJ^{1}+K\wedge e^{2}.
\end{equation*}%
Since $J^{2}=0$, one has 
\begin{equation*}
DJ^{1}=dJ^{1}=\varepsilon \pi e^{-t}\cos (\pi y)\,dx\wedge dy,
\end{equation*}%
and 
\begin{equation*}
K\wedge e^{2}=-a_{0}e^{-t}\cos (\pi y)\,dx\wedge dy.
\end{equation*}%
Therefore, 
\begin{equation*}
\partial _{t}T^{1}=(\varepsilon \pi -a_{0})e^{-t}\cos (\pi y)\,dx\wedge dy.
\end{equation*}%
This decomposition makes explicit the separate roles of frame evolution and
connection evolution in torsion transport \cite{Yavari2012}\cite%
{Srinivasa2021}\cite{Fumeron2022}

\paragraph{Interpretation.}

Torsion evolves through two distinct geometric mechanisms: 
\begin{equation*}
\partial_t T^1 = DJ^1 + K \wedge e^2,
\end{equation*}
where $DJ^1$ represents material frame evolution and $K \wedge e^2$
represents connection evolution.

\subsubsection{Scalar Reduction and Engineering Interpretation}

In two dimensions, write 
\begin{equation*}
T^1 = \tau(x,y,t)\,dx \wedge dy, \qquad \Omega = \kappa(x,y,t)\,dx \wedge dy.
\end{equation*}

For Example 1, 
\begin{equation*}
\tau = e^{-t}\sin(\pi y), \qquad \kappa = -\pi e^{-t}\cos(\pi y).
\end{equation*}

For Example 2, 
\begin{equation*}
\tau = e^{-t}(a_0-\varepsilon \pi)\cos(\pi y), \qquad \kappa = a_0 \pi
e^{-t}\sin(\pi y).
\end{equation*}

The constitutive laws reduce to 
\begin{equation*}
H_1 = \mu_T \tau, \qquad O = \mu_R \kappa.
\end{equation*}

The force stress may be written as 
\begin{equation*}
\Sigma _{1}=\sigma _{y}\,dy,\qquad \sigma _{y}=\partial _{y}H_{1}.
\end{equation*}

\bigskip

\subsection{Analytical Example with Numerical Evaluation of Dynamic Bianchi
Transport in 2D}

This appendix provides an illustrative evaluation of the two-dimensional
reduction of the mesoscopic Cosserat system. The goal is to visualize how
defect transport governed by the dynamic Bianchi identities produces
configurational forces.

\subsubsection{Kinematic setup}

We consider a planar domain with coframe 
\begin{equation*}
e^1 = dx, \qquad e^2 = dy,
\end{equation*}
and a scalar connection 
\begin{equation*}
\omega = a(y,t)\,dx, \qquad a(y,t)=e^{-t}\sin(\pi y).
\end{equation*}

\subsubsection{Defect fields}

\begin{equation*}
T^1 = a(y,t)\,dx \wedge dy, \qquad T^2 = 0,
\end{equation*}
\begin{equation*}
\Omega = -a_y(y,t)\,dx \wedge dy.
\end{equation*}

\subsubsection{Dynamic Bianchi identities}

\begin{equation*}
K = \partial_t \omega = a_t(y,t)\,dx,
\end{equation*}
\begin{equation*}
\partial_t T^1 = K \wedge e^2, \qquad \partial_t \Omega = dK.
\end{equation*}

These relations verify explicitly that the dynamic Bianchi identities govern
the evolution of torsion and curvature fields.

\subsubsection{Configurational force density}

For the translation generator $E=\partial_x$, and using the configurational
force expression derived in the main text, we obtain 
\begin{equation*}
F_{\partial_x}(y,t) = \frac{\pi}{2} e^{-2t}\sin(2\pi y)\,dx \wedge dy,
\end{equation*}
for a representative choice of constitutive parameters.

\subsubsection{Numerical evaluation}

At time $t=0.5$, selected values are: 
\begin{equation*}
\begin{array}{c|c|c|c}
y & a(y,t) & \Omega (y,t) & F_{\partial _{x}}(y,t) \\ \hline
0.25 & 0.4289 & -1.3484 & 0.6736 \\ 
0.50 & 0.6065 & 0 & 0 \\ 
0.75 & 0.4289 & 1.3484 & -0.6736%
\end{array}%
\end{equation*}

\subsubsection{Numerical Illustration of Dynamic Bianchi Transport}

This section illustrates the configurational force density $F(y,t)$
associated with defect transport governed by the dynamic Bianchi identities.

\paragraph{Multiple-time profiles}

\begin{figure}[h]
\centering \includegraphics[width=0.7\textwidth]{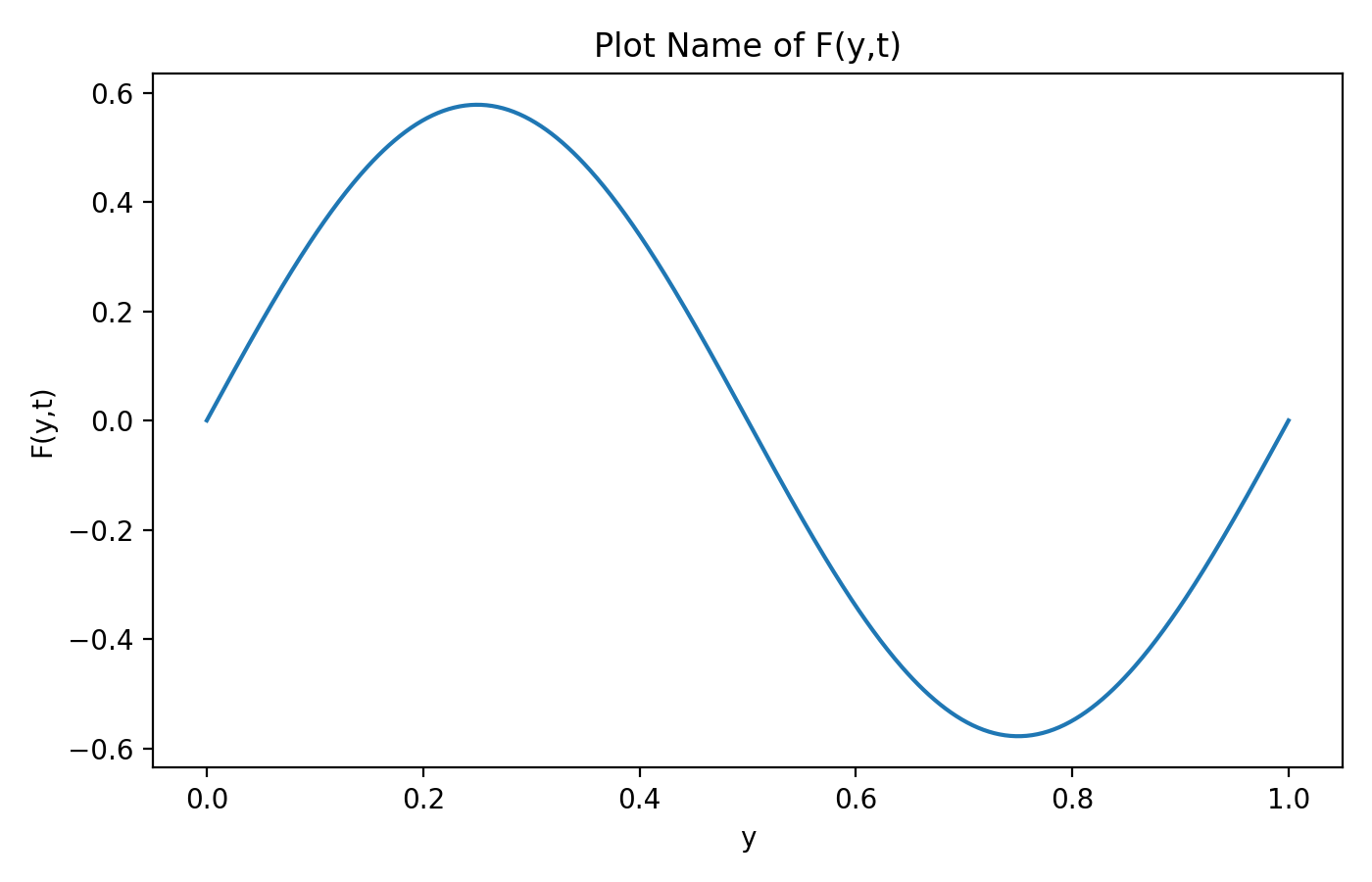}
\caption{Configurational force density $F(y,t)$ at different times ($t=0,
0.5, 1$). The amplitude decays exponentially while preserving spatial
structure.}
\end{figure}

\paragraph{Spatiotemporal evolution}

\begin{figure}[h]
\centering \includegraphics[width=0.7\textwidth]{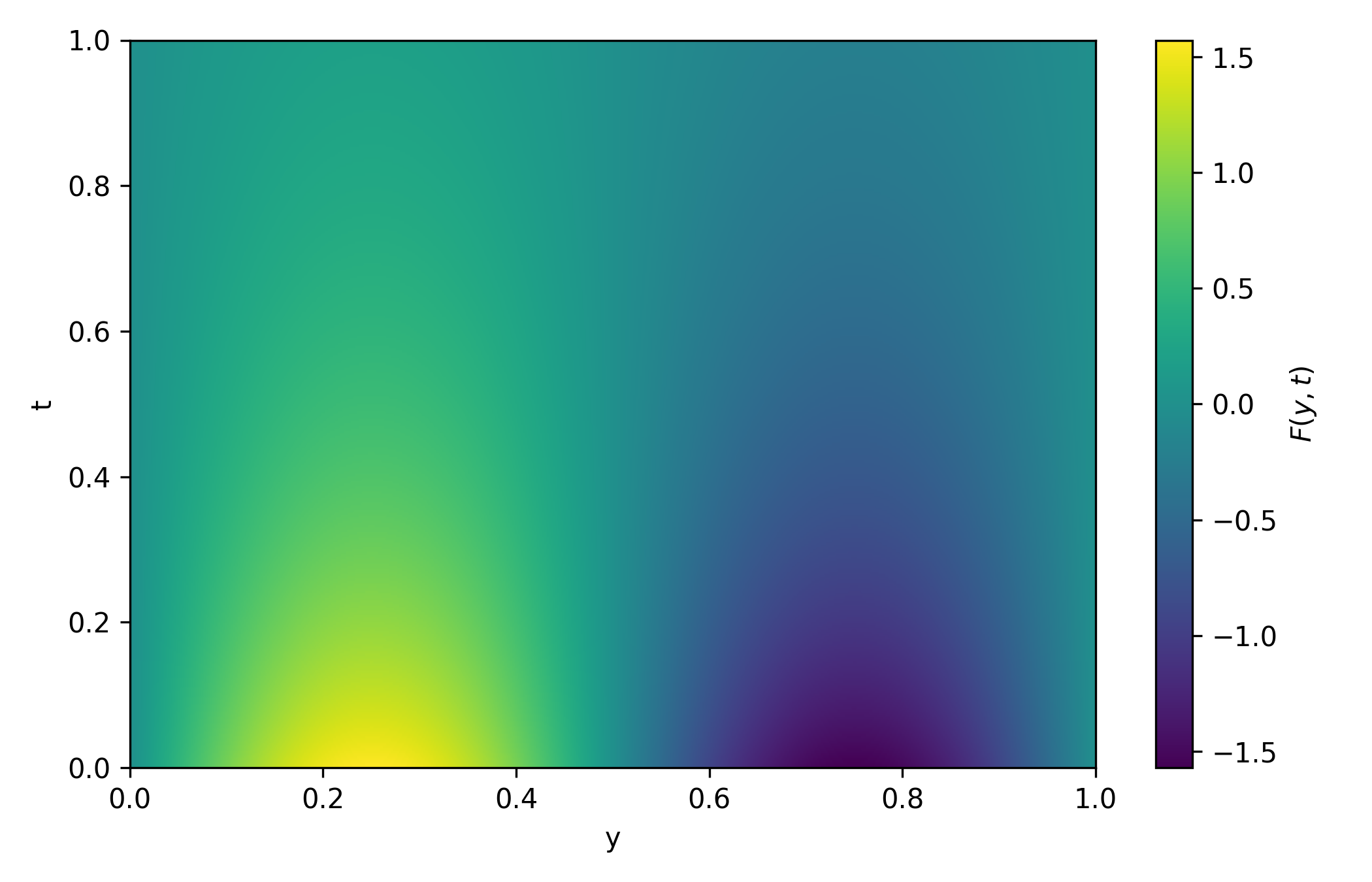}
\caption{Spatiotemporal evolution of $F(y,t)$ showing exponential decay in
time and oscillatory spatial structure.}
\label{Spatio}
\end{figure}

\subsubsection{Interpretation}

The results show that:

\begin{itemize}
\item the rotation field generates both torsion and curvature,

\item curvature changes sign across the domain,

\item the configurational force vanishes at symmetry points,

\item the force changes sign, indicating opposite directions of defect
motion,

\item the magnitude decays in time proportionally to $e^{-2t}$.
\end{itemize}

This behavior is consistent with defect transport in structured continua,
where gradients of curvature generate configurational forces driving defect
motion.

\subsubsection{Conclusion}

This example demonstrates that the dynamic Bianchi identities act as
transport laws for defect fields, and that configurational forces arise
directly from this transport. The structure parallels Maxwell-type evolution
while retaining the geometric coupling characteristic of Cosserat media.

\subsection*{Summary}

These examples demonstrate that:

\begin{itemize}
\item defect transport is governed by the dynamic Bianchi identities \cite%
{Hehl1995} \cite{Roychowdhury2013},

\item excitation and stress fields follow from constitutive relations,

\item configurational forces  \cite{Eshelby1951}\cite{Maugin1993} emerge as
responses to defect transport governed by the Bianchi identities\cite{Hehl1995} \cite{Roychowdhury2013},

\item both coframe and connection dynamics contribute to incompatibility.
\end{itemize}

\section*{Conclusion}

We have developed a mesoscopic extension of Cosserat elasticity motivated by
the breakdown of compatibility in the classical theory. Once
compatibility-breaking perturbations are admitted, the classical formulation
ceases to remain self-consistent under admissible variations, necessitating
an enlargement of the constitutive framework. This naturally leads to a
mesoscopic description in which torsion and curvature are treated as
independent distributed measures of defects, consistent with the geometric
theory of defects developed by Kondo (1952) \cite{Kondo1952}and Kr\~{A}\P %
ner (1981)\cite{Kroner1981}.

Within this setting, the coframe and connection are introduced as
independent fields in a Palatini-type variational formulation, in the spirit
of metric-affine theories (Hehl et al., 1995)\cite{Hehl1995}. The resulting
Euler--Lagrange equations recover the standard balance laws while
simultaneously introducing defect-related excitation fields. Material
invariance yields configurational force and moment balances, whose kinematic
structure is governed by the dynamic Bianchi identities .\cite{Hehl1995} 
\cite{Roychowdhury2013} As a result, configurational forces \cite%
{Eshelby1951}\cite{Maugin1993} and moments arise intrinsically from the
variational structure rather than being postulated separately, extending the
configurational mechanics framework of Eshelby (1951)\cite{Eshelby1951} and
Maugin (1993) \cite{Maugin1993}.

The present framework extends classical configurational mechanics by
providing a fully covariant formulation in terms of differential forms,
incorporating distributed torsion and curvature as primary variables, and
unifying translational and rotational configurational balances \cite{Maugin1993} \cite{Gurtin2000}through a Noether-based approach. It also connects to
generalized continuum theories such as micromorphic and gradient elasticity
models (Forest, 2009)\cite{Forest2009}, as well as gauge-theoretic
descriptions of defects (Lazar, 2010)\cite{Lazas2010}, while replacing singular
defect measures with smoothly distributed fields embedded in a variational
setting. Related geometric approaches to defects in structured media further
support this viewpoint (Roychowdhury, 2017; Fumeron, 2022).\cite%
{Roychowdhury2017}\cite{Fumeron2022}

A quadratic constitutive model together with its dissipative extension
demonstrates that the theory accommodates both reversible and irreversible
processes. The resulting structure exhibits a Maxwell-type analogy and
offers a unified framework for defect transport, configurational
interactions, and microstructural evolution.

The key contribution of this work is thus the formulation of a variationally
consistent and geometrically covariant mesoscopic theory of Cosserat media
with distributed defects. This framework provides a foundation for future
developments, including the analysis of localization phenomena, defect
concentration, and computational implementations for structured solids with
evolving internal geometry.

\end{document}